\documentclass[useAMS]{mn2e}

\usepackage{graphicx}
\input{epsf}

\def\be{\begin{equation}}
\def\ee{\end{equation}}
\def\ba{\begin{eqnarray}}
\def\ea{\end{eqnarray}}
\def\de{\partial}
\def\12{{1\over 2}}

\def\ltsima{$\; \buildrel < \over \sim \;$}
\def\simlt{\lower.5ex\hbox{\ltsima}}
\def\gtsima{$\; \buildrel > \over \sim \;$}
\def\simgt{\lower.5ex\hbox{\gtsima}}

\begin{document}

\title[Particle decay and 21 cm from the early universe]{\bf Particle 
decay in the early universe:
predictions for 21 cm}
\author[Yu. A. Shchekinov and E. O. Vasiliev]
       {Yu. A. Shchekinov$^1$\thanks{E-mail:yus@phys.rsu.ru}
and E. O. Vasiliev$^{2,3}$\thanks{E-mail:eugstar@mail.ru} \\
$^1$Department of Physics, University of Rostov, 
Sorge St. 5, Rostov-on-Don, 344090 Russia\\
$^2$Tartu Observatory, 61602 T\~oravere, Estonia\\
$^3$Institute of Physics, University of Rostov, 
Stachki Ave. 194, Rostov-on-Don, 344090 Russia
}
\date{Accepted 2006 December 15.
      Received 2006 April 1;
      in original form 2006 April 1}
\pagerange{\pageref{firstpage}--\pageref{lastpage}}
\pubyear{2006}
\maketitle

\label{firstpage}

\begin{abstract}
The influence of ultra-high energy cosmic rays (UHECRs) and decaying 
dark matter particles on the emission and absorption characteristics 
of neutral hydrogen in 21 cm at redshifts $z = 10-50$ is considered. 
In presence of UHECRs 21 cm can be seen in absorption 
with the brightness temperature $T_b=-(5-10)$ mK in the 
range $z=10-30$. Decayng particles can stimulate a 21 cm 
signal in emission with $T_b\sim 50-60$~mK at $z =50$, and 
$T_b \simeq 10$~mK at $z \sim 20$. Characteristics of the 
fluctuations of the brightness temperature, in particular, its  
power spectrum are also calculated. The maps 
of the power spectrum of the brightness 
temperature on the plane {\it wavenumber-redshift} are 
shown to be sensitive to the parameters of UHECRs and decaying 
dark matter. 
Observational possibilities to detect manifestations of 
UHECRs and/or decaying particles in 21 cm with the future radio telescopes 
(LOFAR, 21CMA and SKA), and to distinguish contributions from them 
are briefly discussed.
\end{abstract}


\section{Introduction}

\noindent

At $z\sim 1000$ the Universe enters the ``dark ages'' epoch, the electrons
and protons recombine, and gas remains neutral until the first
luminous objects emerge at $z\sim 30-20$. Neutral gas can be observed
in a redshifted 21 cm line of neutral hydrogen in emission or absorption
against the cosmic microwave background (CMB). This gives a possibility
for studying the processes associated with transition of the neutral
universe into a fully ionized state, and for identification of 
the sources responsible for reionization (Hogan \& Rees 1979, 
Madau et al 1997). At present observations of 21 cm line are 
anticipated as a promising tool for diagnostics of the universe in the 
end of ``dark ages'' (Madau et al 1997, Tozzi et al 2000,
Ciardi \& Madau 2003).

Possible sources of photons which can be important for reionization 
form two different groups. The first, connected 
with conventional baryon physics, involves 
the baryons processed in stellar nucleosynthesis (Shapiro \& Giroux 1987,
Miralda-Escud\'e \& Rees 1994, Tegmark et al 1994, Cen 2003, Ciardi et al 2003, 
Choudhury \& Ferrara 2006) and in shock
waves near black holes (Madau et al 1999, Oh 2001, Ricotti \& Ostriker 2004). 
The other is connected with unstable dark matter and can 
include massive neutrinos (Sciama 1982, Scott et 
al 1991), superheavy X-particles (Berezinsky et al 1997, 
Kuzmin \& Rubakov, 1998, Birkel \& Sarkar, 1998). 
The first luminous objects are commonly thought to be the principal source
of the reionization. They heat gas in the universe through ionization by
ultraviolet (UV) and X-ray photons. This inevitably affects the emissivity
of gas in the 21 cm line, because the hydrogen spin temperature $T_s$
depends on the gas kinetic temperature $T_k$, and thus observed intensity
imprints the effects from the objects of the first generation (Madau et al 1997,
Tozzi et al 2000, Ciardi \& Madau 2003, Loeb \& Zaldarriaga 2004,
Zaldarriaga et al 2004, Chen \& Miralda-Escud\`e 2004, Sethi 2005).

Production of copious number of
ionizing photons during the dark ages can be connected also 
with the origin of ultra-high energy cosmic ray (UHECRs) if they form from
decaying superheavy dark matter (SHDM) particles with masses
$M_X\simgt 10^{12}$~GeV in the so-called top-to-bottom scenario
(Berezinsky et al 1997, Kuzmin \& Rubakov, 1998, Birkel \&
Sarkar, 1998). The associated production of UV photons can have strong
influence on cosmological recombination (Doroshkevich \& Naselsky 2002, Doroshkevich et al 2003). 

Decaying dark matter particles, such as massive neutrinos, can also
contribute significantly to reionization (Sciama 1982,
Dodelson \& Jubas 1994). Initial polarisation measurements of the CMB by
the Wilkinson Microwave Anisotropy Probe (WMAP) sattelite
(Spergel et al 2003) imparted a new impulse to this possibility.
The obtained relatively large optical depth of the universe
$\tau\simeq 0.16$ suggested unrealistically strong constraints
on properties of the first stellar objects (e.g. Cen 2003, Wyithe \& 
Loeb 2003, see also an alternative discussion in Tumlinson et al 
2004), which lead some to consider that decaying dark matter can be at least a complementary source of reionization (Hansen \& Haiman 2004,
Chen \& Kamionkowski 2004 (hereafter CK), 
Kasuya et al 2004, Kasuya \& Kawasaki 2004,
Pierpaoli 2004, Mapelli et al 2006). The corresponding heating 
can change characteristics of 21 cm in emission and absorption. 
Very recently, the influence of dark matter decay and annihilation 
on the 21 cm line from dark ages has been also considered by 
Furlanetto et al. (2006a) for long living particles.  For 
the models with similar parameters of decaying particles they 
reached conclusions close to ours. 

In this paper we study the effects of UHECRs and decaying particles 
on cosmological 21 cm background. We assume a $\Lambda$CDM 
cosmology with the parameters 
$(\Omega_0, \Omega_{\Lambda},\Omega_m,\Omega_b,h)=(1.0,0.76,0.24,0.041,0.73)$ 
(Spergel et al 2006).

\section{Spin temperature}

The two processes: atomic collisions and scattering of UV photons, couple
the HI spin temperature and the gas kinetic temperature 
(Wouthuysen 1952, Field 1958)

\be
T_s = {T_{cmb} + y_a T_k + y_c T_k \over 1 + y_a + y_c}
\ee
here $T_{cmb}$ is the CMB temperature, $y_c, y_a$ are the functions 
determined by the collisional excitations and the intensity of the UV 
resonant photons

\be
y_a = {P_{10}T_\ast \over A_{10} T_k}, \hspace{1cm}
y_c = {C_{10}T_\ast \over A_{10} T_k}
\ee
$T_\ast = 0.0682~K$ is the hyperfine energy splitting,
$A_{10} = 2.87\times 10^{-15}$~s$^{-1}$ is the spontaneous emission
rate of the hyperfine transition,  $C_{10} = k_{10}n_H + \gamma_e n_e$
is the collisional de-excitation rate by hydrogen atoms and electrons,
the rate by protons is negligible, for $k_{10}$ we use the approximation
by Kuhlen et al (2006),
for $\gamma_e$ we take the approximation from Liszt (2001),
$P_{10}$ is the indirect
de-excitation rate, which is related to the total Ly$\alpha$ scattering
rate $P_a$ (Field 1958)

\be
P_{10} = 4 P_a / 27,
\ee
where

\be
P_a = \int c n_\nu \sigma_\nu d\nu,
\ee
$n_\nu$ is the number density of photons 
per unit frequency range, $\sigma(\nu)$ is the
cross section for Ly$\alpha$ scattering (Madau et al 1997).
The brightness temperature in 21 cm is then (Field 1958, Chen \&
Miralda-Escud\'e 2004)

\be
T_b = 25{\rm mK} {T_s-T_{cmb}\over T_s}
            \left(\Omega_b h_0 \over 0.03\right)
            \left(0.3 \over \Omega_{m0}\right)^{1/2}
            \left(1+z\over 10\right)^{1/2}.
\ee
It is worth mentioning here that recently Hirata \& Sigurdson (2007) showed that the spin temperature depends in general on atomic velocities $T_s=T_s(v)$, if excitation and de-excitation collisional processes are dominated by interatomic (H-H and H-He) 
collisions. This effect results in a sufficiently 
strong (up to 60\%) increase of the 21-cm line width, however 
only in a 2\% decrease of the emissivity. 
In this paper we focus mostly on the total emissivity, and 
therefore neglect these effects. In principle, such a widening 
of the 21 cm line 
can be important in the power spectrum of the brightness 
temperature on very small scales, $\lambda\simlt 3$ kpc.


\section{Ionization and thermal history of the universe}

Evolution of the fractional ionization is described by

\be
{dx_e \over dz} = {1\over (1+z) H(z)}(R_s(z) - I_s(z) - I_e(z))
\label{xe}
\ee
where besides the standard recombination and ionization rates $R_s(z), I_s(z)$,
the term $I_e(z)$ is explicitly introduced due to presence of additional
sources of UV photons. For the UHECRs this term can be written in the 
form (Peebles et al 2000, Doroshkevich \& Naselsky 2002)

\be
I_e(z) = \epsilon(z) H(z) x_{\rm H},
\ee
where $\epsilon(z)=\epsilon_0/(1+z)$ is the production of ionizing
photons, $H(z)$ is the Hubble parameter, 
$x_{\rm H}$ is the fraction of neutral hydrogen. For decaying particles
this term is written as (CK)

\be
I_e(z) = \chi_i f_x \Gamma_X { m_p c^2 \over h \nu_c}
\ee
where $\chi_i$ is the energy fraction deposited into ionization for which we 
made use the calculations by Shull \& van Steenberg (1985), 
$m_p$ is the proton mass, $f_x =\Omega_X(z)/\Omega_b(z)$, $\Omega_b(z)$ is 
the baryon density parameter, $\Omega_X(z)$ is the fractional abundance of 
decaying particles, $\Gamma_X$ is the decay rate, $h\nu_c$ is the energy 
of Ly-c photons. For short living particles 
$\Omega_X(z)\propto e^{-\Gamma_Xt}$ and their contribution to ionization and 
heating essentially vanishes at $t>\Gamma_X^{-1}$; normally $\Gamma_X^{-1}$ 
is assumed to be comparable to the comoving Hubble time in the range of 
redshifts of interest. 

The gas temperature is determined by equation

\ba
(1+z){dT\over dz} = 2T + k_C {x_e \over H(z) (1+f_{He}+x_e)} (T-T_{cmb}) \\
\nonumber
                       - {2\over 3k}{K \over H(z) (1+f_{He}+x_e)}
\label{temp}
\ea
where the second term in the r.h.s. describes the energy exchange between
the gas and CMB photons $k_C =  4.91\times 10^{-22} T_{CMB}^4$, $f_{He} = 0.24$,
the third term is the heating from additional sources of the 
ionizing photons, $K$ is the corresponding heating rate which can be 
written in the form (CK) 

\be
K = \chi_h m_p c^2 f_x \Gamma_X
\ee
where $\chi_h$ is the energy fraction depositing into heating; as for 
$\chi_i$ we used for $\chi_h$ the results of Shull \& van Steenberg (1985). 
By order of magnitude $\chi_i\sim\chi_h\sim 1/3$ for the conditions 
we are interested in. 

In the presence of UHECRs the final products 
are only Ly$\alpha$ and Ly-c photons (Doroshkevich \& Naselsky 2002). 
As pointed out by (Chen \& Miralda-Escud\'e 2004) the injected Ly$\alpha$ photons
change the gas temperature very little: $K_\alpha = 0$. Ly-c photons have 
the energy slightly in excess of the hydrogen binding energy: 
$\sim$ 13.6 eV+$3kT/2$. Therefore the heating rate 
from the UHECRs produced Ly-c photons $K_c$ has to be calculated 
self-consistently together with equation (\ref{temp}). However, as we will 
see later the gas kinetic temperature 
in presence of UHECRs is at most 150 K, so that the heating rate from 
Ly-c photons is always less than $\sim 2\times 10^{-31}\epsilon(1+z)^{3/2}$ 
erg cm$^{-3}$ s$^{-1}$. The corresponding 
contribution to the gas kinetic temperature is less than a few percent. 

We apply a modified version of the code RECFAST (Seager et al 1999) to solve
equations (\ref{xe}), (\ref{temp}).

\section{Results}

\noindent

Stellar and quasi-stellar sources of ionizing radiation
begin to form at redshifts $z<20$, and gas around them heated 
through photoionization can emit in 21 cm with a spot-like distribution.
The spots have too small angular sizes, and can be detected only
later, when at $z\simlt 15$ star formation increases and forms 
pervading domains of sufficiently hot gas 
(Zaroubi \& Silk 2004). Instead, UV photons from 
additional ionizing sources, like UHECRs and decaying particles,
illuminate and heat the IGM homogeneously, and therefore the signal from 
21 cm can be more easily detected from the earliest redshifts. 

\begin{figure}
\vspace{34pt}
\includegraphics[width=76mm]{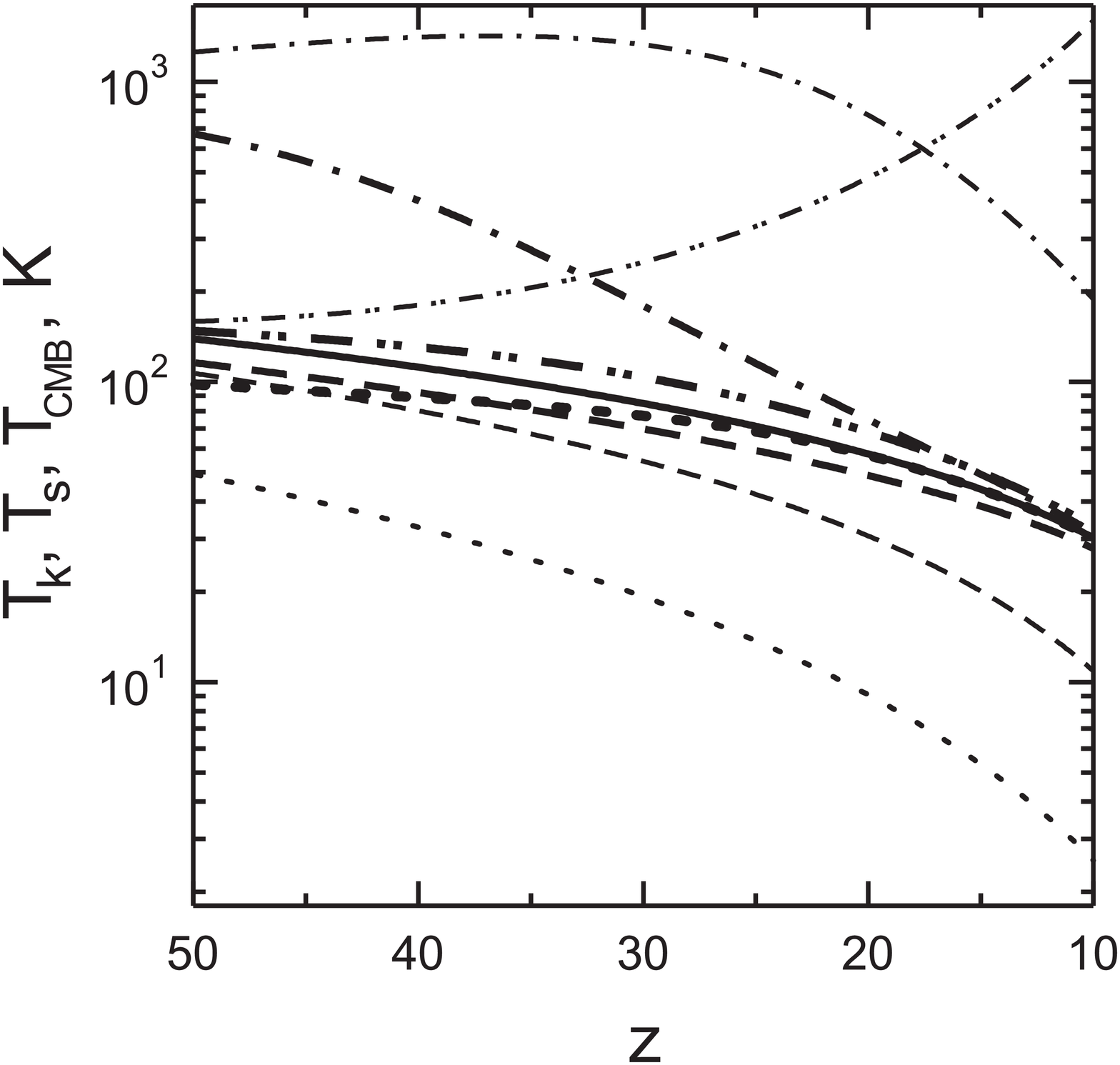}
\vspace{12pt}
\caption{The kinetic (thin lines), spin (thick lines) and CMB (thick solid line) 
temperatures for the standard recombination $\epsilon = 0$ (dashed),
in the presence of UHECRs for the UV production rate $\epsilon = $ 1 (dotted),  
and in the presence of decaying particles:
long living with $\xi = 3\times 10^{-26}$~s$^{-1}$ (dot-dot dashed), 
and short living with $\Gamma_X = 10^{-15}$~s$^{-1}$, $f_X(z_{eq}) = 10^{-8}$ 
(dash-dotted).  
}
\label{fig1}
\end{figure}

\begin{figure}
\vspace{16pt}
\includegraphics[width=85mm]{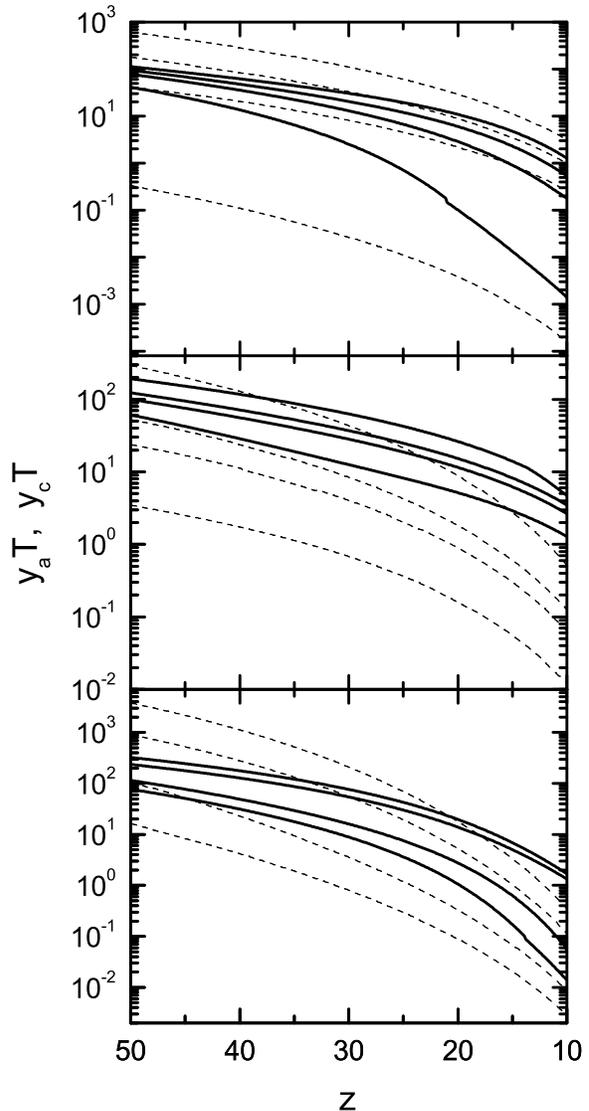}
\caption{
The contributions to the spin temperature from 
collisions (thick solid lines) and photon exitations 
(thin dashed lines) for the models
with UHECRs {\it upper panel} for $\epsilon = 0,~0.3,~1,~3$ 
from bottom to top; long living particles {\it mid panel}: 
$\xi = 6\times 10^{-27}$~s$^{-1}$, 
 $3\times 10^{-26}$~s$^{-1}$, 
 $6\times 10^{-26}$~s$^{-1}$, 
 $3\times 10^{-25}$~s$^{-1}$ -- from the lowermost to the uppermost; 
$\xi = \chi_i f_X \Gamma_X$, 
and short living {\it lower panel} decaying particles: 
$\Gamma_X = 10^{-14}$~s$^{-1}$, $f_X(z_{eq}) = 0.5\times 10^{-8}$, 
$\Gamma_X = 5\times 10^{-15}$~s$^{-1}$, $f_X(z_{eq}) = 10^{-8}$, 
$\Gamma_X = 10^{-15}$~s$^{-1}$, $f_X(z_{eq}) = 10^{-8}$,
$\Gamma_X = 10^{-15}$~s$^{-1}$, $f_X(z_{eq}) = 5\times 10^{-8}$ -- 
from the lowermost to the uppermost.
}
\label{fig2}
\end{figure}

Fig.~1 shows kinetic and spin temperatures for a selected set of 
ionizing photon production rates by UHECRs and decaying particles. 
For the standard history ($\epsilon = 0$) the spin
temperature is nearly equal to the CMB value at $z\leq 20$, and the
intergalactic gas cannot produce distinguishible signal in 21 cm in this
redshift range. The kinetic
temperature for UHECRs with $\epsilon = 1$ is more than twice of
the value in the standard ionization history ($\epsilon = 0$). The increase 
of kinetic temperature for higher $\epsilon$ is due to an increase of 
the fractional ionization of gas and a stronger coupling between the CMB photons 
and electrons. As a result, the gas kinetic temperature in this case shows similar 
variation with $z$ as the CMB temperature. 
Due to collisional de-excitations of the hyperfine structure level 
this increase in $T_k$ unavoidably results in a decrease
of the spin temperature, such that the difference between the 
spin and CMB temperature in the range $z \sim 35-10$ is for 
$\epsilon=1$ greater than for $\epsilon = 0$; $T_s-T_{CMB}$ vanishes 
at $z\leq 10$. This means that the signal in 21 cm 
can be detected in the redshift range $z\sim 10-35$ in absorption. 
The long living particles provide a permanent heating 
the injection energy rate $3k\dot T/2= K/(1+f_{\rm He}+x_e)$, such that 
for a sufficiently high heating (ionization) rate shown in Fig. 1
the gas kinetic temperature in this case grows towards lower $z$ 
as seen in Fig. 1 (thin dot-dot dashed line). Contrary, the short living particles 
inject heat with the rate $\dot T\propto e^{-\Gamma_Xt}$, which manifests 
in a relatively fast decrease of the kinetic temperature at low $z$ 
(thin dash-dotted line in Fig. 1). The HI spin temperature remains in 
both cases above the CMB temperature.

As mentioned above, the heating of gas in models with UHECRs is small, and therefore the deviation of the brightness temperature from the standard value 
is mostly due to the Wouthuysen-Field (WF) effect from the UV 
photons produced by UHECRs. In Fig. 2 we show the contributions of heating (collisional excitation and de-excitation) and 
Wouthuysen-Field effect to the spin temperature for UHECRs 
and decaying particles.

\begin{figure}
\vspace{24pt}
\includegraphics[width=85mm]{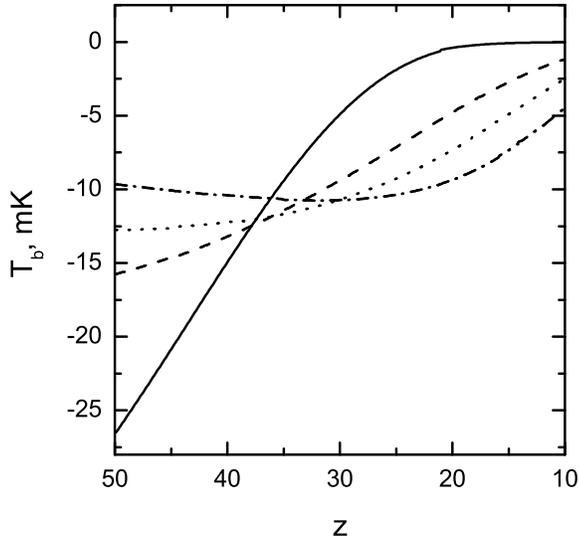}
\caption{The brightness temperature in the presence of UHECRs for
the UV production rate $\epsilon = $ 0 (solid), $\epsilon=0.3$ (dashed), 
$\epsilon=1$ (dotted), $\epsilon=3$ (dash-dotted).
}
\label{fig3}
\end{figure}

Fig. 3 presents brightness temperature versus redshift for several values
of $\epsilon$. An obvious qualitative difference between the standard model
($\epsilon=0$) and the models with $\epsilon\geq 0.3$ is clearly seen:
contrary to the standard case in all models $T_b$ flattens at $z>30$
at the level $T_b\simeq -(10-15)$ mK. On the other hand at $z<25$, where
the standard model shows almost zero $T_b$ all models with $\epsilon\geq 0.3$
have brightness temperature between -5 and -10 mK.  
The signal of tens mK 
can though be easily swamped by a much stronger (of $\simgt 100$ K) 
foreground emission at meter wavelengths. However, 
multi-frequency observations 
seem to allow one to remove the foreground, which is expected to be 
featureless in frequency space (Shaver et al. 1999, 
Di Matteo et al. 2002, Oh \& Mack 
2003, Gnedin \& Shaver 2004, Zaldarragia et al. 2004). One can 
hope thus that an 1000 hours
LOFAR (Low Frequency Array) and/or SKA (Square Kilometre Array) 
observation can discriminate between the standard and
$\epsilon\geq 0.3$ models (see discussion in Sect. 4).

\begin{figure}
\vspace{24pt}
\includegraphics[width=85mm]{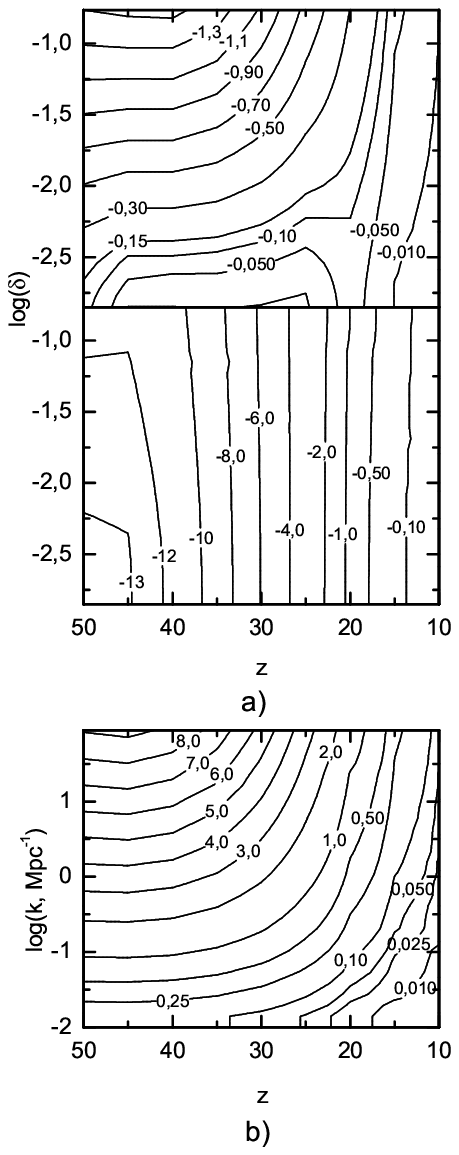}
\caption{ a) The difference between the local 
and all-sky brightness 
temperatures $\delta T_b=T_b(\delta)-T_b$ (upper panel), the derivative $dT_b/d\delta$ 
(lower panel) as functions of redshift and density variation $\delta$; 
b) the temperature power spectrum $\delta T_b(k,z)$ as a function of redshift 
and wavenumber $k$: standard model. Numbers on isocontours are given in mK. 
}
\label{fig4}
\end{figure}

\begin{figure}
\vspace{24pt}
\includegraphics[width=85mm]{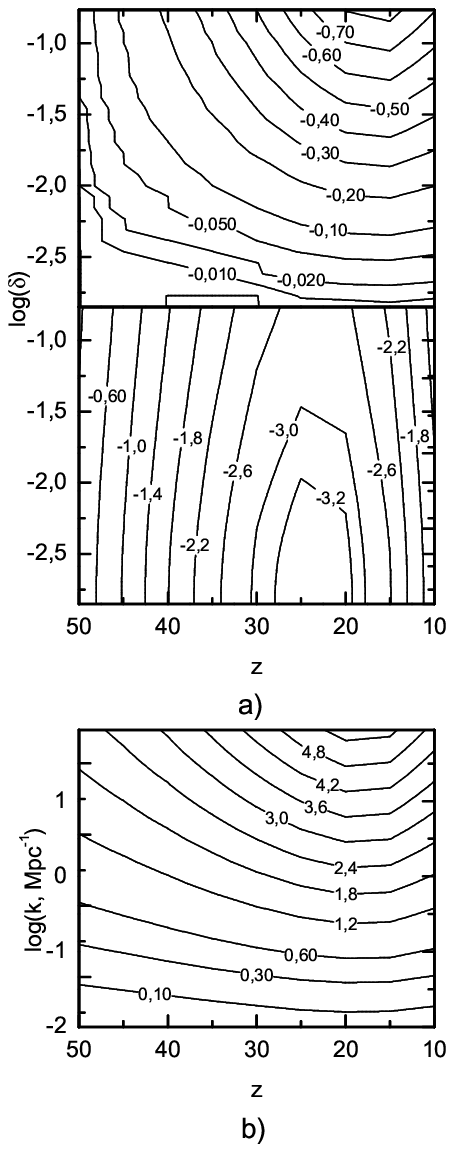}
\caption{Same as in Fig. 4 for model with UHECRs: $\epsilon=1$.
}
\label{fig5}
\end{figure}

Spatial variations of the 
brightness temperature connected with density perturbations in 
emitting gas are to be more easily distinguished in the 
redshifted 21 cm emission. For this purpose 
we calculated the derivative of the brightness temperature $T_b$ 
over the amplitude of the density fluctuation $\delta$, where the 
perturbed density field is assumed in the form 
$\rho=\rho_0(z)(1+\delta)$, with $\rho_0(z)$ being the background 
baryon density. We calculated the deviations of the spin and 
kinetic temperatures starting from $z_0=1000$ 
and followed the evolution of perturbed regions according to 
equation (\ref{temp}) to the redshift of 
interest $z$, accounting the dependence of kinetic coefficients 
$k_{10}$ and $\gamma_e$ on temperature. In Fig. 4 we 
show results for the standard recombination
model: the upper panel of Fig. 4a illustrates 
the deviation $\delta T_b$ of the local brightness temperature 
from the all-sky one for a separate perturbation of a given 
density amplitude 
$\delta$. As expected, 
the overdense regions show larger amplitude in absorption. The 
dependence of $\delta T_b$ on redshift ($x$-axis) weakens at 
higher $z$. For the sake of clearity in the lower panel of Fig. 4a 
we show also the derivative $d T_b/d\delta$ versus redshift and 
amplitude: at $z=20-40$ it weakly 
depends on the amplitude $\delta$, although changes very sharp versus 
redshift. The total signal comprises contributions from the whole 
spectrum of perturbations and is represented by the temperature power 
spectrum (Barkana \& Loeb 2005ab, Hirata \& Sigurdson 2007) 
\be
P_{T_b}=P_{\mu^0}(k)+\mu^2P_{\mu^2}(k)+\mu^4P_{\mu^4}(k), 
\ee
where $\mu=k_{||}/k$ is the cosine of the angle between the 
line of sight and the wavevector, 
\begin{eqnarray}
&&P_{\mu^0}=\left({\de T_b\over \de \delta}\right)^2P_{\delta}(k),
\nonumber\\
&&P_{\mu^2}=k \bar T_b{\de T_b\over \de \delta}P_{\delta,v}(k), \\
&&P_{\mu^4}=k^2\bar T_b^2P_{v}(k),
\nonumber
\end{eqnarray}
$\bar T_b$ is the mean brightness temperature, $P_{\delta}(k)$, 
$P_v(k)$ are the power spectrum of the baryon 
density and velocity fluctuations, $P_{\delta,v}(k)$, their 
cross-spectrum. Fig. 4b shows $\delta T_b(k,z)
=\sqrt{k^3P_{T_b}/2\pi^2}$ 
for the standard model in the plane $(z,k)$: 
the dominant contribution stems from the density fluctuations, 
the contribution from density-velocity cross-spectrum is always 
around 20-30 \%, the velocity fluctuations contribute less 
than 10-20 \% at $z\leq 40$, and increases (up to 30\%) at 
$z\sim 50$; these estimates are given for the average (over 4$\pi$) 
value of $\mu^2=1/3$. 

In Fig. 5 we show the distributions $\delta T_b$, $dT_b/d\delta$ and 
$\delta T_b(k,z)$ for the model with UHECRs $\epsilon=1$. As in the 
standard model 21 cm line can be observed in absorption, however 
there is a clearly seen distinct feature between the two cases: 
the standard model shows gradient of the temperature power spectrum 
from low amplitudes and redshifts (right bottom corner in 
Fig. 4b) towards higher amplitudes and redshifts (left upper 
corner), while the model with UHECRs reveals an opposite behaviour 
of the gradient of $\delta T_b(k,z)$ -- from the left lower to the 
right upper corner, and in addition, a much smaller gradient in 
redshift (i.e. in frequencies) at lower amplitudes. 

\begin{figure}
\vspace{24pt}
\includegraphics[width=85mm]{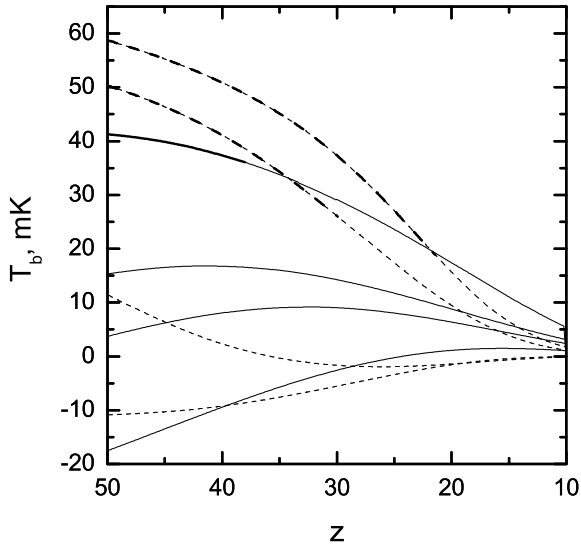}
\caption{ {\it Solid lines} -- the brightness temperature in the presence of 
long living particles $\Gamma_X \ll H_0$:   
$\xi = 6\times 10^{-27}$~s$^{-1}$, 
 $3\times 10^{-26}$~s$^{-1}$, 
 $6\times 10^{-26}$~s$^{-1}$, 
 $3\times 10^{-25}$~s$^{-1}$ -- from the lowermost to the uppermost; 
$\xi = \chi_i f_X \Gamma_X$. 
{\it Dashed lines} -- the brightness temperature in the 
presence of short 
living particles $\Gamma_X \simgt H_0$: 
$\Gamma_X = 10^{-14}$~s$^{-1}$, $f_X(z_{eq}) = 0.5\times 10^{-8}$, 
$\Gamma_X = 5\times 10^{-15}$~s$^{-1}$, $f_X(z_{eq}) = 10^{-8}$, 
$\Gamma_X = 10^{-15}$~s$^{-1}$, $f_X(z_{eq}) = 10^{-8}$,
$\Gamma_X = 10^{-15}$~s$^{-1}$, $f_X(z_{eq}) = 5\times 10^{-8}$ -- 
from the lowermost to the uppermost; the thick part of the solid 
curve and the dot-dashed thick parts of the dashed curves indicate 
the predominance of WF effect. 
}
\label{fig6}
\end{figure}

The models with decaying dark matter can produce 21 cm   
in emission and in absorption as well. 
Fig.~6 shows the all-sky brightness temperature for the 
long living (solid lines) and short living (dashed lines) decaying 
particles. The gas kinetic temperature grows 
with heating rate and can be lower or greater than $T_{cmb}$ depending 
on $\xi$. Therefore, at low heating rate ($\xi\leq 6\times 10^{-27}$ s$^{-1}$) 
gas can be observed in 21 cm only in absorption 
with a lower absolute brightness temperatue than in the standard 
case, essentially approaching it due to decrease 
in the heating rate (compare the lower solid
line in Fig. 6 with the
solid line in Fig. 3). The transition
from absorption to emission occurs at $\xi\geq 10^{-26}$ s$^{-1}$. 
The long living decaying particles affect the 
spin and the brightness temperature mostly through collisional 
heating (as seen in Fig. 2), however in some cases Wouthuysen-Field effect can become more important: this is  
the model with $\zeta=3\times 10^{-25}$ s$^{-1}$ at $z>40$ -- 
the contribution from WF effect is shown in Fig. 6 as a thick
part of the corresponding curve. 

Brightness temperature for short living particles 
$T_b$ increases with the heating rate 
similar to what occurs for the long living particles, and correspondingly 
at a sufficiently high heating 21 cm can be observed in emission. There is, 
however, a qualitative difference between the dependence of brightness 
temperature on redshift in these two cases: for the long living particles 
$T_b(z)$ has always negative second derivative, while in the case of the 
short living particles $T_b(z)$ has an inflection 
point. Note in this connection that $T_b(z)$ curves in models with the UHECRs 
have always positive second derivative. Similarly to 
the models with long living decaying particles, for short living particles a mostly collisional (heating) contribution to 
$T_s$ and $T_b$ is 
reversed to predominantly Wouthuysen-Field effect for the 
models with $\Gamma_X=10^{-15}$ s$^{-1}$ and $f_X(z_{eq})=
10^{-8}$, and with $\Gamma_X=10^{-15}$ s$^{-1}$,  
$f_X(z_{eq})=5\times 10^{-8}$ at $z>30$ and $z>20$, respectively; this is also shown by the thick parts of the corresponding curves in 
Fig.  6. The inflection point on $T_b(z)$ nearly 
corresponds to redshifts where WF effect becomes dominating. 

The differences between the second derivatives of the all-sky 
brightness temperature $T_b(z)$ is reflected to a certain extent on 
the behaviour of the temperature power spectra on $(z,k)$ plane. As 
seen in Fig. 7b the lines of equal spectral power for 
the long living particles reveal a smooth, nearly uniform, gradient 
over the whole range of redshifts and wavenumbers (amplitudes). At the same time, for the short living particles (Fig. 8b) 
these lines thicken at the low $z$ side, with a stronger 
gradient, similarly to the model with UHECRs where $d^2T_b/dz^2$ is always positive. The behaviour of the temperature difference 
$\delta T_b$ and its derivative $d T_b/d\delta$ on the 
$(z,k)$ plane are also distinct for these two models of 
decaying particles (Fig. 7a and 8a): the short 
living particles show thickening near the low end of redshifts. 
These differences though are not well pronounced 
(as, for instance, between the standard model and the model with UHECRs shown in Fig. 4 and 5), and it seems to be challenging to reveal them. In the high redshift end $z>20$ (lower frequences), 
the $\delta T_b(k,z)$ maps for the two models look quite similar.  
However, at $z<20$ (higher freqiences) the temperature power 
spectrum at a given wavenumber decreases towards lower redshift 
faster in the model with short living particles, which can be 
recognizable in the frequency space. 

\begin{figure}
\vspace{24pt}
\includegraphics[width=85mm]{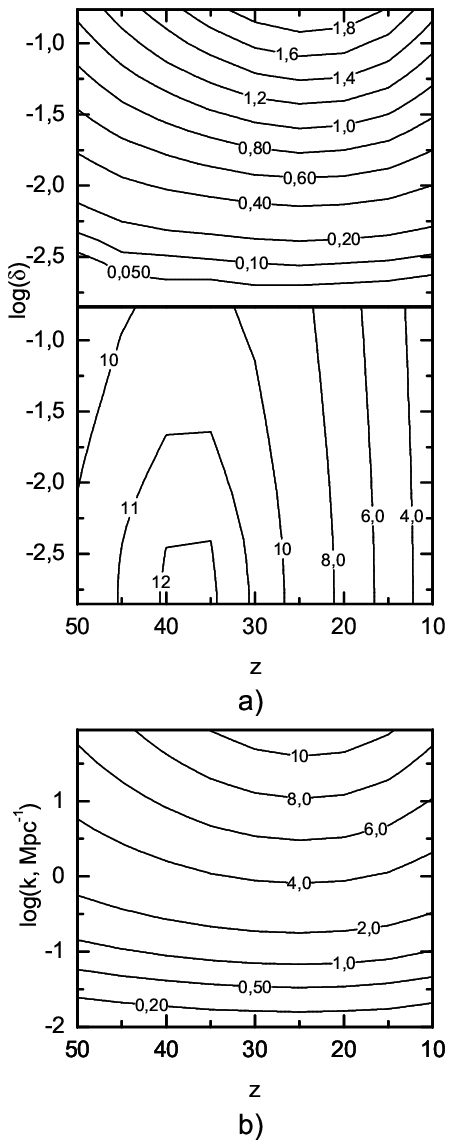}
\caption{Same as in Fig. 4 for the model with 
long living 
particles $\Gamma_X \ll H_0$ for  
$\xi =  6\times 10^{-26}$~s$^{-1}$.  
}
\label{fig7}
\end{figure}

\begin{figure}
\vspace{24pt}
\includegraphics[width=85mm]{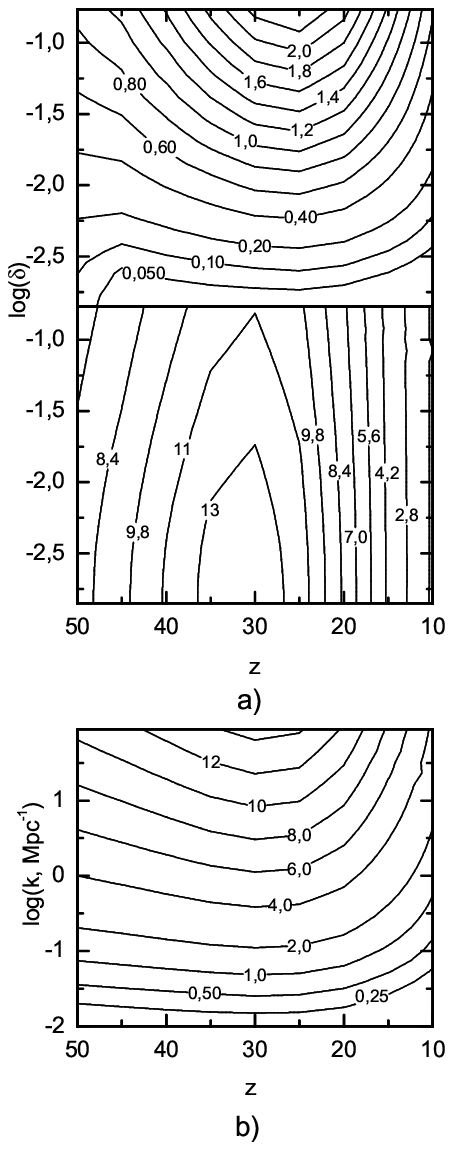}
\caption{Same as in Fig. 4 for the model with 
 short 
living particles $\Gamma_X \simgt H_0$: 
$\Gamma_X = 10^{-15}$~s$^{-1}$, $f_X(z_{eq}) = 10^{-8}$.
}
\label{fig8}
\end{figure}

The differences between the second derivatives of the all-sky 
brightness temperature over redshift $d^2T_b/dz^2$ for the models 
with UHECRs and long- and short living particles, or the 
corresponding differences between the temperature power spectra, 
imprint in spectral features of the 21 cm line. This circumstance may have a principal 
significance for choosing a strategy for observational discrimination between 
manifestations from the three sources of the ionizing photons, 
provided that observations are possible for a wide range of redshifts, 
from $z=10$ to $z=50$. However, if the qualitative difference between 
$T_b(z)$ and $\delta T_b(k,z)$ curves for the UHECRs and long living particles seems to be 
relatively easily observationally distinguishible, it looks problematic 
for the curves $T_b(z)$ and $\delta T_b(k,z)$ in cases with the long and short living particles 
because the inflection of $T_b(z)$ in the latter case lies within 
$\Delta T_b(z)\simeq 5$ mK, and the differences between the 
$\delta T_b(k,z)$ maps are basically within a few mK for wavenumbers 
from, e.g., 0.1 to 1 Mpc$^{-1}$. From this point of view 
``two-color'' diagrams, connecting the differences between 
the brightness temperatures at different wavelengths, can be complementary 
to the analysis of $T_b(z)$ curves  
and $\delta T_b(k,z)$ maps. We show an example of such ``two-color'' 
diagrams for the three sources of UV photons: UHECRs, long and short living 
decaying particles in Fig.~9. Specifically, we plot the relative all-sky temperature differences  
$\Delta^2 T_{23}=\Delta T_b(\lambda_2)-\Delta T_b(\lambda_3)$ versus 
$\Delta^2 T_{12}=\Delta T_b(\lambda_1)-\Delta T_b(\lambda_2)$, where 
$\lambda_1=210$ cm, $\lambda_2=420$ cm, $\lambda_3=840$ cm, for a set of 
models for these three cases, here $\Delta T_b(\lambda_i)$ is 
the difference 
between the total signal at $\lambda_i=21(1+z_i)$ cm and the 
temperature of the foreground emission at this wavelength. Since 
the foreground emission has temperature of several orders of magnitude higher than the expected from 21 cm, determinantion of the 21 
cm signal with using such a substraction procedure 
seems challenging. However, 
as pointed out by Di Matteo et al. (2002), and Oh \& Mack 
(2003), the prospectives for measurements of the all-sky signal from 
redshifted 21 cm are not totally dim because the foreground 
emission is expected to be featureless in frequency (see also discussion in Zaldarriaga et al. 2004, Sethi 2005). 
The three cases fall into quite distinct regions 
in the ``two-color'' plane with a separation of $\Delta^2 T>5$ mK, and thus can in 
principle be discriminated observationally. 

\begin{figure}
\vspace{24pt}
\includegraphics[width=85mm]{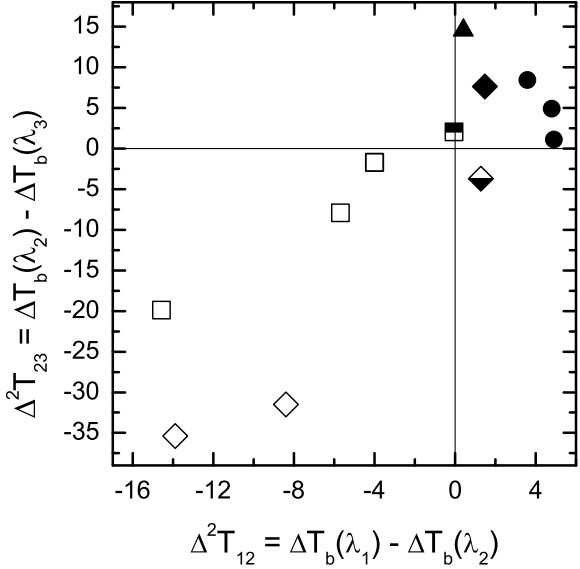}
\caption{``Two-color'' diagram for the models with UHECRs and decaying 
particles. Open symbols show differences in brightness temperatures in 
emission, filled -- in absorption; half-filled correspond to emission 
at low $z$ (lower half is open) and absorption at high $z$ (upper half 
is filled), and vise versa. Circles show $(\Delta^2 T_{12},\Delta^2 T_{23})$ 
for UHECRs, diamonds -- for long living decaying particles, squares -- 
for short living particles, the triangle depicts the standard 
(no additional ionizing photons) case. 
}
\label{fig9}
\end{figure}

The parameters of decaying particles and UHECRs 
we explored in this work are consistent with the current constraints. 
CK found using WMAP data $\xi<10^{-24}$~s$^{-1}$ for the long living 
particles, and a slightly weaker upper limit for the short living ones. 
Recent results from WMAP give a lower value of the optical depth $\tau\simlt 0.09$ 
(Spergel et al 2006). This strengthens the constrains on the parameters 
of decaying particles compared to those inferred from the WMAP data of the  
first year. However, as pointed out by CK decaying particles with 
short lifetimes do not affect singnificantly the optical
depth and are thus less constrained. On the other hand, the upper limit 
for $\xi=3\times 10^{-25}$ s$^{-1}$ for long living particles, we consider here, 
lies within the restrictions on  possible ionization rates 
corresponding to the optical depth obtained from the third year WMAP data.  
For UHECRs Doroshkevich \& Naselsky (2002) inferred from the MAXIMA-1 and BOOMERANG data (de Bernardis et al 2000, Hanany et al 2000) 
$\epsilon\leq 3$. Observations of HI at $z=10-50$ in 21 cm can provide 
further constraints. 

The sensitivity of ongoing long-wavelength experiments, 
such as LOFAR, 21CMA (former PAST), MWA, LWA, 
SKA and LUDAR (Carilli 2006) seems to be 
sufficient to detect signal in 21 cm affected by decaying particles and UHECRs at pre-ionization epochs. 
The minimum background source flux density reqiured to detect an
absorption feature with the signal to noise ratio $S/N$ is 
(Furlanetto 2006) 
\begin{eqnarray}
&&S_{min} = 16\left({S/N \over 5}\right)\left({10^{-3}
\over \tau}\right)\left({10^6{\rm m^2}\over A_{eff}}\right)
 \nonumber\\
&& \times
 \left({T_{sys}\over 400~{\rm K}}\right)
 \left({100~{\rm kHz}\over \Delta\nu}~{1~{\rm week}\over t_i}\right)^{0.5}
 ~{\rm mJy}
\end{eqnarray}
where $\tau$ is the optical depth in 21 cm, $A_{eff}$ is
the effective area of telescope, $\Delta\nu$ is the bandwidth of
channel, $t_i$ is the total integration time; in the low frequency 
($\sim 100$ MHz) range the system 
temperature $T_{sys}$ is dominated by the Galactic synchrotron
background and scales as $\nu^{-2.5}$, or in terms of redshift 
(Chen \& Miralda-Escud\`e 2006, Furlanetto et al. 2006b) 
\be
T_{sys} \simeq 2000 \left({1+z \over 21}\right)^{2.5}~{\rm K}. 
\ee
For a conservative estimate we take the bandwidth of a channel 
$\Delta\nu = 4$~MHz, as a typical
value for future telescopes. The effective area of the telescopes 
varies from $\sim 3\times 10^4$m$^2$ for LOFAR (at 75~MHz) 
to $10^6$~m$^2$ 
for SKA and LWA (for LUDAR it can reach up to $\sim 10^7$~m$^2$). 
The optical depth in 21 cm increase from 
$10^{-3}$ at $z\simeq 10$ to 0.01 at $z\ge 15$ (Furlanetto 2006). 
Therefore, for the redshifts $z\ge 10$ we are interested in $\tau$ 
can be taken $>10^{-3}$. 
The integration time is assumed to be of 10 weeks.
Thus, the minimum background flux is 12~mJy at $z = 20-40$ 
for LOFAR, 0.4~mJy at $z = 20$ for SKA 
and LWA, and 2.3~mJy at $z = 40$ for LWA. In the case of UHECRs
the difference of a smooth signal expected between the models 
with $\epsilon \ge 0.3-3$ is $\ge 5$~mJy, and is even larger in 
the models with decaying dark matter. This supports the conclusion 
that observations on future telescopes can discriminate between 
the standard model and those affected by UHECRs and decaying 
particles.

\section{Conclusions}

In this paper we have considered the influence of UV photons from 
decaying dark matter particles and UHECRs on the ability of neutral hydrogen 
to emit and/or absorb in 21 cm at redshifts $z = 10-50$. We have found that

\begin{itemize}
\item the three sources of additional ionizing photons: long 
living and short living unstable dark matter particles and 
UHECRs produce fairly distinct dependences of the all-sky 
brightness temperature on redshift $T_b(z)$ -- the first and 
the third give negative and positive second derivatives of the 
curves $T_b(z)$, while the second has $T_b(z)$ with an 
inflection point. Although the all-sky 21 cm emission is 
expected to be swamped by the owerwhelming foreground signal, 
one can hope that due to the lack of features in frequency space 
the latter can be removed. Moreover, the maps of the power spectrum 
of the brightness temperature on the wavenumber-redshift plane, 
reveal clear differences (seen in gradients of the temperature 
power spectrum over redshift) between various models of the 
particles. 

\item these features manifest in the frequency 
space of 21 cm line. From this point of view three wave-band observations at $\lambda_1$, $\lambda_2$ and
$\lambda_3$ and a ``two-color'' diagram for the relative 
(with respect to the foreground emission at a given wavelength) 
all-sky temperature differences 
$\Delta^2 T_{23}=\Delta T_b(\lambda_2)-\Delta T_b(\lambda_3)$
versus
$\Delta^2 T_{12}=\Delta T_b(\lambda_1)-
\Delta T_b(\lambda_2)$ can provide an additional
tool for discrimination between the sources of ionizing photons 
in the end of dark ages.

\end{itemize}

In general, decaying dark matter particles can have strong effects
on overall history of the universe: they may change its thermal evolution
and cosmological nucleosynthesis
(Scherrer 1984, Vayner et al 1985,
Vayner \& Shchekinov 1986), dynamics of
large scale structure formation
(Doroshkevich et al 1989,
Bharadwaj \& Sethi 1998,
Cen 2001),
reionization regime (Sciama 1982, Dodelson \& Jubas 1994,
Hansen \& Haiman 2004,
Chen \& Kamionkowski 2004,
Kasuya et al 2004,
Kasuya \& Kawasaki 2004,
Pierpaoli 2004, Mapelli et al 2006), and even formation of the 
fisrt stars though enhancement of H$_2$ molecule formation 
(Shchekinov \& Vasiliev 2004, Vasiliev \& Shchekinov 2006, 
Biermann \& Kusenko 2006, Ripamonti et al 2007,
Furlanetto et al. 2006a). 
21 cm emission from the epochs ending the dark ages can carry the
imprints from decaying particles, and seems a promising tool for
understanding
their properties.
Future radio telescopes (such as LOFAR, 21CMA, MWA, LWA and SKA)
seem to have sufficient flux sensitivity for detection the
signal in 21 cm influenced by decaying particles and UHECRs.

\section{Acknowledgements}

We thank the anonymous referee for valuable criticism. 
This work is supported by the Federal Agency of Education 
(project code RNP 2.1.1.3483) 
and by the RFBR (project code 06-02-16819-a).



\end{document}